\documentclass[12pt]{article}
\usepackage{amsmath,amssymb,graphicx,mathrsfs,hyperref}

\newcommand{\be}{\begin{equation}}
\newcommand{\ee}{\end{equation}}
\newcommand{\bea}{\begin{eqnarray}}
\newcommand{\eea}{\end{eqnarray}}

\def\({\left(} \def\){\right)}

\begin{document}

\title{\vspace{-1.8in}
\vspace{3mm}
\vspace{0.3cm} The two canonical conjugate pairs at the horizon of a D1D5 black hole}
\author{\large  Merav Hadad   \\ Levy Rosenblum \\
\vbox{\begin{flushleft} {\small
\hspace{.2in} Department of Natural Sciences, The Open University of Israel P.O.B. 808, Raanana 43107, Israel.
 E-mail:  meravha@openu.ac.il ; levyros@gmail.com}\vspace{7pt}\\
  \end{flushleft}}
 }
\date{}
\maketitle

\begin{abstract}

 The Euclidian opening angle at the $r-t_E$ surface, $\Theta_{r-t_E}$ at the horizon of a black hole is canonically conjugate to the black hole entropy.  We prove that for a $D1D5$ black hole there exists in addition to this pair, another canonical pair: the opening angle at the $r-y$ surface, $\Theta_{r-y}$ and a Wald like term $S_{Wr-y}$. This leads to an uncertainty at $\Theta_{r-y}$ which suggests that the surface $r-y$ is actually a superposition of surfaces with different conical singularities. This corresponds to the same type of singularities obtained by string theory excitations of a $D1D5$ black hole.
\end{abstract}



\newpage

Black hole entropy is expected to help reveal aspects of quantum gravity.  Geometrical properties of the black hole phase space, dictated by canonical relations at the horizon, may help identify some characteristics of the gravitational degrees of freedom.

The Euclidean opening angle $\Theta_{r-t_E}$ in the $r-t_E$ surface at the horizon of a black hole is canonically conjugate to its entropy \cite{Carlip:1993sa} and thus we can expect the uncertainty condition $\Delta\Theta_{r-t_E}\Delta S\geq\hbar$. Since the black hole entropy is a Noether charge \cite{Wald:1993nt,Iyer:1994ys}, this property can be extended to general theories of gravity as well \cite{Brustein:2012sa}.  Though this uncertainty is a result of fluctuations which are expected in a theory of quantum gravity, the fact that a canonical conjugate pair defines a phase space means that this pair may also be the characteristics of the degrees of freedom of some geometric/gravitational microstates. Moreover, since uncertainty at the opening angle corresponds to a superposition of metrics with different conical singularities, we should expect that whatever the microstates that create gravity (loops/triangles/strings/...), they must have the ability to form metrics with different conical singularities. This property of the microstates of gravity seems to be natural for discrete theories of gravity \cite{Dittrich:2008va}, and it was also found in string theories.

 String theory succeeds in describing a kind of quantum gravity, and with its aid one can identify the entropy of a BPS black hole as string theory excitations \cite{Strominger:1996sh,Mathur:2005zp}. Some of these vibrations create metrics with quantized conical singularities in the $r-y$ surface \cite{Giusto:2012yz}, and thus we expect that the black hole metric should be a kind of superposition of metrics, some of which have a conical singularity in the $r-y$ surface. This creates an uncertainty at the opening angle at the $r-y$ surface. However, from the canonical relation of $\Theta_{r-t_E}$ and the black hole entropy we would expect an uncertainty at the opening angle at the $r-t_E$ surface and not at the $r-y$ surface.

In this paper we show that for a $D1D5$ black hole there exists in addition to the known canonical pair, $[\Theta_{r-t_E},S_W]$, another canonical pair: the opening angle at the $r-y$ surface: $\Theta_{r-y}$ and a Wald like term $S_{Wr-y}$. This leads to an uncertainty at $\Theta_{r-y}$ which predicts that the surface $r-y$ is actually a superposition of surfaces with conical singularities. This corresponds to the type of singularity obtained from string theory excitations and thus suggests that the uncertainty at  $S_{Wr-y}$, $\Theta_{r-t_E}$  and the black hole entropy may also come from some other type of string vibration.

This paper is organized as follows:
We begin by dividing space into a two dimensional surface and a $D-2$ dimensional submanifold. Then we identify the canonical term of the extrinsic curvature for the Einstein-Hilbert Lagrangian, find a condition for the number of canonical pairs being minimal, and discuss the extension of this to general theories of gravity. Next we check the condition for a spherically symmetric metric, and show that when the two dimensional surface is the $r-t$ surface the condition is fulfilled only on the black hole horizon, which leads to the uncertainty between the Euclidean opening angle in the $r-t_E$ surface and the Wald entropy, as was found in \cite{Brustein:2012sa}. Since for $D1D5$ metrics the condition for the minimal canonical pair is not fulfilled, we extend the method by dividing space into a $1+n$ and a $D-1-n$ dimensional submanifolds, and find an additional condition on the manifold. Finally, we show that for $D1D5$ black hole metric the new condition is fulfilled for $n=2$ and thus there exists, in addition to the known canonical pair, $[\Theta_{r-t_E},S_W]$, another canonical pair: the opening angle at the $r-y$ surface: $\Theta_{r-y}$ and a Wald like term $S_{Wr-y}$. We discuss the implication of this result on the problem of finding the missing microstates of the $D1D5$ black hole.

We begin with a given vector field $n_a$, which is not a geodesic so that $n^b\nabla_b n_a=a u_a$ where $a(x^i)\neq0$ is a scalar function of the coordinates \footnote{Thus, if $n_a$ is time like, $a$ is the magnitude of the acceleration of an observer who moves along $n_a$.}, and $u_a$ is a unit vector normal to $n_a$. Next we define some scalar function  $t(x^i)$ such that $t^a\nabla_at=1$ where  $t_a=Nn_a$ (as in general $t_a=Nn_a+V_a$ where $N$ and $V_a$ are the lapse function and the shift vector respectively, but one can always use a coordinate transformation which will change the metric so that $V_a=0$ at least locally. In the examples we use in this paper the metrics are orthogonal and the vector field $n_a$ is along one of the axis so that $V_a=0$ everywhere.) Note that in this metric the interval along $n_a$ is $ds=\sqrt{dx_idx^i}=Ndt$.  Next we foliate the space time with respect to the unit vector field $u_a$ by defining  $(D-1)$-hypersurfaces which are normal to $u_a$. The lapse function $M$ and shift vector $W_a$ satisfy $r_a=Mu_a+W_a$ where $r^a\nabla_ar=1$ and $r$ is constant on $\Sigma_{D-1}$.  The $\Sigma_{D-1}$ hypersurfaces metric $h_{ab}$ is given by $g_{ab} =h_{ab}+(-1)^{s_u}u_a u_b$, where $s_u=0$ if $u_a$ is space-like and $s_u=1$ if $u_a$ is time-like.
The extrinsic curvature of the hypersurfaces is given by $K_{ab}=-\frac{1}{2}\mathcal{L}_u h_{ab}$ where $\mathcal{L}_u$ is the Lie derivative along $u^a$.
The $D-2$ hypersurfaces defined by $t=const$ and $r=const$ are thus normal to the given vector $n_a$ and to $u_a$.  The D-2 hypersurfaces metric $\sigma_{ab}$ is given by $h_{ab} =\sigma_{ab}+(-1)^{s_n}n_a n_b$, where $s_n=0$ if $n_a$ is space-like and $s_n=1$ if $n_a$ is time-like. \footnote{In addition $n_a$ and $u_a$ must fulfill Frobenius's theorem. All the $n_a$ and $u_a$ used in the examples on this paper fulfill this theorem.}

Next we write the Einstein-Hilbert Lagrangian $\int d^Dx\sqrt{-g}R$ where $R=g^{ac}g^{bd}{^{(D)}R_{abcd}}$, in terms of the two vectors $u_a$,$n_a$ and quantities related to the $D-2$ hypersurfaces, as follows: We use
\begin{eqnarray}
\label{canonical}
^DR_{abcd}= {^{D-1}R_{abcd}}-4u_{[a}\mathcal{L}_uK_{b][c}u_{d]}+ f(u_a,M,D_aD_bM,K_{ad},D_aK_{bc})
\end{eqnarray}
where $D_a$ denotes the covariant derivative of the $r=const$ hypersurfaces,
 $\mathcal{L}_r=M\mathcal{L}_u+\mathcal{L}_W$ and $\sqrt{-g}=M\sqrt{-h(-1)^{s_u}}=MN\sqrt{-\sigma(-1)^{s_u+s_n}}$. The Lagrangian becomes:
\begin{eqnarray}
\label{canonical1}
L=\int M dr N dt d^{D-2}x\sqrt{-\sigma(-1)^{s_u+s_n}}g^{ac}g^{bd} {^{D-1}R_{abcd}}\nonumber \\
-4\int dr N dt d^{D-2}x\sqrt{-\sigma(-1)^{s_u+s_n}}g^{ac}g^{bd}u_{[a}\mathcal{L}_r( K_{b][c})u_{d]}\nonumber \\
+4\int dr N dt d^{D-2}x\sqrt{-\sigma(-1)^{s_u+s_n}}g^{ac}g^{bd}u_{[a}\mathcal{L}_WK_{b][c}u_{d]}\nonumber \\
+\int M dr N dt d^{D-2}x\sqrt{-\sigma(-1)^{s_u+s_n}}g^{ac}g^{bd}f(u_a,M,D_a D_b M,K_{ad},D_a K_{bc}).
\end{eqnarray}
We can write $K_{bc}$ as $\sigma_b^m\sigma_c^nK_{mn}+2\sigma_b^mn_cn^nK_{mn}+n_bn^mn_cn^nK_{mn}$ and the second line becomes:
\begin{eqnarray}
\label{canonical en}
-4\int dr N dt d^{D-2}x\sqrt{-\sigma(-1)^{s_u+s_n}}g^{ac}g^{bd}u_{[a}\mathcal{L}_r(K_{b][c})u_{d]}=\\\nonumber
-4\int dr N dt d^{D-2}x\sqrt{-\sigma(-1)^{s_u+s_n}}g^{ac}g^{bd}\left\{u_{[a}n_{b]}n_{[c}u_{d]}\mathcal{L}_r(K_{nn})+K_{nn} u_{[a}\mathcal{L}_r(n_{b]} n_{[c})u_{d]}\right\}\\\nonumber
+\int dr N dt d^{D-2}x\sqrt{-\sigma(-1)^{s_u+s_n}}g^{ac}g^{bd}F(\sigma_c^nK_{mn})
\end{eqnarray}
where $K_{nn}\equiv n^b n^a K_{ab}$, and $F(\sigma_c^nK_{mn})$ involves terms which depend linearly on the Lie derivative of the projection of the extrinsic curvature on the $\sum_{D-2}$ hypersurface.
Thus $K_{nn}$ is canonically conjugate to $4N\sqrt{-\sigma(-1)^{s_u+s_n}}g^{ac}g^{bd} u_{[a}n_{b]} n_{[c}u_{d]}$ and $n_{b}n_{c}$ is canonically conjugate to $4N\sqrt{-\sigma(-1)^{s_u+s_n}}g^{ac}g^{bd} K u_{a}u_{d}$. \footnote{Since $h_{ab}=(-1)^{sn}n_{a}n_{b}+\sigma_{ab}$,  the term $4\sqrt{-h(-1)^{s_u}}Ku^{b}u^{c}$ is actually a contribution to the canonical conjugate to $n_{b}n_{c}$.}

 One can also obtain the canonical term of $\sigma_b^m\sigma_c^nK_{mn}$ and $\sigma_b^mn_cn^nK_{mn}$ from the term  $F(\sigma_c^nK_{mn})$. In this paper we try to find the minimal number number of phase space of the extrinsic curvature, thus we focus on the cases where the components of the extrinsic curvature on the $D-2$ submaniold vanish: $\sigma_c^nK_{mn}\simeq0$. This condition seems to be sufficient for the black holes we are dealing with in this paper.

However, in order to establish the same canonical relation that appears in \cite{Carlip:1993sa}, we need to identify a slightly different canonical pair. The canonical pair should be $[K_{nn},4\sqrt{-\sigma(-1)^{s_u+s_n}}g^{ac}g^{bd} u_{[a}n_{b]} n_{[c}u_{d]}]$, which is without the shift $N$. This is actually the canonical pair as seen by an observer moving on a trajectory along $n_a$. To see this we define the "proper" interval along $n_a$: $d\tau =Ndt$. Thus if we write (\ref{canonical en}) as seen by an observer moving along $n_a$ it becomes:
\begin{eqnarray}
\label{canonicalend}
4\int dr d\tau d^{D-2}x\sqrt{-\sigma(-1)^{s_u+s_n}}g^{ac}g^{bd}u_{[a}\mathcal{L}_r(K_{b][c})u_{d]}=\\\nonumber
4\int dr d\tau d^{D-2}x\sqrt{-\sigma(-1)^{s_u+s_n}}g^{ac}g^{bd}\left\{u_{[a}n_{b]}n_{[c}u_{d]}\mathcal{L}_r(K_{nn})+K_{nn} u_{[a}\mathcal{L}_r(n_{b]} n_{[c})u_{d]}\right\}\\\nonumber
+\int dr d\tau d^{D-2}x\sqrt{-\sigma(-1)^{s_u+s_n}}g^{ac}g^{bd}F(\sigma_c^nK_{mn})
\end{eqnarray}
and $K_{nn}$ will be canonically conjugate to $4\sqrt{-\sigma(-1)^{s_u+s_n}}g^{ac}g^{bd} u_{[a}n_{b]} n_{[c}u_{d]}$  for an observer who moves on a trajectory along $n_a$.

 In the quantum limit this becomes
\begin{eqnarray}
\label{canonical relation}
\left[K_{nn}(x),4\sqrt{-\sigma(-1)^{s_u+s_n}} g^{ac}g^{bd} u_{[a}n_{b]} n_{[c}u_{d]}(\tilde{x})\right]=\hbar \delta^{D-1}(x-\tilde{x}).
\end{eqnarray}

Actually, it seems that this may be generalized to general theories of gravity. For this purpose we need to use Brown's work \cite{Brown:1995su}, where he found that in general theories of gravity the canonical conjugate variable of the extrinsic curvature $K_{bc}$ is  $\sqrt{-h(-1)^{s_u}}u_a u_d U_0^{abcd}$ (where $U_0^{abcd}$ is an auxiliary variable which equals $\frac{\partial\mathscr{L}}{\partial R_{abcd}}$ when the equations of motion hold). Thus by replacing $4g^{ac}g^{bd}$ with $U_0^{abcd}$ and using the same steps as before one finds that $K_{nn}$ is canonically conjugate to $\sqrt{-\sigma(-1)^{s_u+s_n}}u_a n_b n_c u_dU_0^{abcd}$.\footnote{Actually the canonical structure of higher derivative theories is much more complicated and involves finding the right field equation by using the Hamiltonian formalism and Dirac brackets (see for example \cite{Avraham:2014twa}). However, as we will see, the possibility that the identification of the extrinsic curvature and its canonical conjugate might be the "real" phase space of general theories of gravity is supported by the fact that near a horizon of a black hole these variables converge with those obtained by other methods which regard the horizon as a boundary.}

In the quantum limit this becomes
\begin{eqnarray}
\label{canonical relation finel}
\left[K_{nn}(x),\sqrt{-\sigma(-1)^{s_u+s_n}} U_0^{abcd}u_a n_b n_c u_d (\tilde{x})\right]=\hbar \delta^{D-1}(x-\tilde{x}).
\end{eqnarray}

If the $D-1$ hypersurface $r=const$ is compact it is tempting to integrate over this compact hypersurface. If $K_{nn}$ does not depend on the coordinates of the $D-2$ submanifold we can do it in two steps: First we integrate on the $D-2$ submanifold: $S_{WL}\equiv\oint u_a n_b n_c u_d\frac{\partial\mathscr{L}}{\partial R_{abcd}}d\Sigma^{D-2}$. This gives us a term that looks like Wald entropy. Second we integrate over  $d\tau$: $\oint K_{nn} d\tau$. Since $d\tau=ds$ where $s$ is an arc along the unit vector $n_a$  and $K_{nn}=u^aa_a\equiv k_g$ which is the geodesic curvature, this  will give us, using the Gauss-Bonnet theorem, the total curvature $\Theta$ (which is  $2\pi$ for any closed space curve\footnote{Two comments: First, when using the Gauss-Bonnet theorem we assume that the Gaussian curvature of the two dimensional surface $n^a-u^a$ can be neglected. This can be done for all the black holes we are dealing with in this paper. Second,  since the Gauss-Bonnet theorem is relevant for Euclidean space, we use the Euclidean limit if $n^a$ or $u^a$ is timelike.}). We end up with the canonical relation and an uncertainty relation:
\begin{eqnarray}
\label{action part2}
\left[\Theta,S_{WL}\right]=\hbar\\
\Delta \Theta \Delta S_{WL}\geq \hbar
\end{eqnarray}
This suggests that the total curvature of a surface defined by a closed orbit along a non-geodesic vector $n_a$ may be canonical to a Wald like term, and if so to fulfill the uncertainty relation and define a phase space.

 As we noted in footnote (4), identifying canonical terms is not that obvious. However if it turns out that these two terms are canonical, this may be an extension of Wald entropy, for an accelerating observer moving along a closed orbit when $n_a$ is timelike and for any non geodesic closed curve if $n_a$ is spacelike.  As we will see, a hint that this may be the case appears when applying this method to a $D1D5$ black hole, but first we start with an ordinary spherically symmetric black hole.

For a spherically symmetric metric and $n_a$ along the time axis, the projection of the extrinsic curvature on $\Sigma_{D-2}$ vanishes at the horizon (Appendix A). In this case the $K_{nn}$ component of the extrinsic curvature will be the only contribution to the phase space. Moreover, the fact that in this case the term $S_{WL}$ turns out to be exactly Wald entropy \cite{Brown:1995su} reinforces the concept of Wald entropy and the opening angle as comprising the gravitational phase space.

 However, for a $D1D5$ black hole the projection of the extrinsic curvature on $\Sigma_{D-2}$ does not vanish at the horizon.

  For a $D1D5$ black hole  we examine the static metric
\begin{eqnarray}
\label{Static spherical}
ds^2=f(r)(-dt^2+dy^2)+f(r)^{-1}(dr^2+r^2d\Omega ^2)+g(r)dz_{i}^{2},
\end{eqnarray}
where $y$ and $z_i$ are compact. We will use the 'naive' geometry \cite{Mathur:2005zp} where: $f(r)=(1+\frac{Q_1}{r^2})^{-1/2}(1+\frac{Q_5}{r^2})^{-1/2}$ and $g(r)=(1+\frac{Q_1}{r^2})^{1/2}(1+\frac{Q_5}{r^2})^{-1/2}$.

Choosing $n^a$ to be in the time direction,
\begin{eqnarray}
\label{Static spherical}
n^a=(f(r)^{-1/2},0,0,0,...,0),
\end{eqnarray}
$u^a$ becomes
\begin{eqnarray}
\label{Static spherical}
u^a=(0,0,f(r)^{1/2},0,...,0),
\end{eqnarray}
and the extrinsic curvature projection on the $D-2$ submanifold becomes
\begin{eqnarray}
\label{extrinsic curvature}
\sigma^{ac}K_{cb}=\left(
      \begin{array}{ccc}
          0 & 0 & 0 \\
           0 & \frac{f'}{2f^{1/2}} & 0 \\
           0 & 0 & 0
       \end{array}
       \right)\bigoplus\left(\frac{-2 f(r)+r f'(r)}{2 r\sqrt{f(r)}}\right)I_{3} \nonumber \bigoplus \left(\sqrt{f(r)}\frac{g'(r)}{2g(r)}\right) I_4
\end{eqnarray}

 In this case, the extrinsic curvature projection on the $D-2$ submanifold does not vanish near the horizon since near the horizon, $\frac{-2 f(r)+r f'(r)}{2 r\sqrt{f(r)}}= \sqrt{f(r)}\frac{g'(r)}{2g(r)}=0$, but $\frac{f'}{2f^{1/2}}\neq0$. Thus for a $D1D5$ black hole we need to extend our formalism by using more unit vectors.

Thus, instead of one unit normal vector $n_a$,  we deal with a special case where we can find $n$ ($n<D-1$) unit vectors: $n_{(i)a}$  ($i=1...n$) which are normal to each other and to the same unit vector $u_a$ which is defined by $n_{(i)}^b\nabla_b n_{(i)a}=a_{(i)} u_a$ and $a_{(i)}\neq0$.  In addition, we demand that in this  special case we are able to define some scalar function of the coordinates $t_{(i)}$ such that $t_{(i)}^a\nabla_at_{(i)}=1$ where  $t_{(i)}^a=N_{(i)}n_{(i)}^a$. In this case  the interval along $n_{(i)a}$ is $ds_{(i)}=\sqrt{dx_idx^i}=N_{(i)}dt_{(i)}$. We define the $D-n-1$ metric $\tilde{\sigma}_{ab}$ as $h_{ab} =\tilde{\sigma}_{ab}+\sum_{i=1}^{n}(-1)^{si}n_{(i)a} n_{(i)b}$. Next, we plug this into eq. (\ref{canonicalend}) and  obtain for an observer who "moves" on a trajectory along $n_{(i)a}$
\begin{eqnarray}
\label{canonical string}
\left[K^{ab}n_{(i)b} n_{(i)a} (x),\sqrt{{\sigma_{(i)}}} U_0^{abcd}u_a n_{(i)b} n_{(i)c}u_d(\tilde{x})\right]=\hbar \delta^{D-1}(x-\tilde{x})
\end{eqnarray}
where $\sqrt{{\sigma_{(i)}}}= \prod_{j\neq i}N_j \sqrt{-\tilde{\sigma}(-1)^{s_u+\sum_{l}sl}}$ (Note that in this case $K^{ab}n_{(i)b} n_{(j)a}$ vanishes for $i\neq j$).

 Finally, we turn back to the $D1D5$ black hole.

 From the similarity between $t$ and $y$ in the Euclidean metric of the $D1D5$ black hole, we find we need two kinds of vectors to fulfill the condition $\tilde{\sigma}^{ab}K_{bc}=0$: $n_{(1)}^a$ which is along the time axis and $n_{(2)}^a$ which is along the compactified axis $y$.

 Indeed, if we choose instead of $n^a$ the two vectors:
\begin{eqnarray}
\label{Static spherical}
n_{(1)}^a=(f(r)^{-1/2},0,0,....,0)\\\nonumber
n_{(2)}^a=(0,f(r)^{-1/2},0,0,....,0)
\end{eqnarray}
and define a $D-3$ matric $\tilde{\sigma}_{ab}$ as $h_{ab} =\tilde{\sigma}_{ab}+(-1)^{s1}n_{(1)a} n_{(1)b}+(-1)^{s(2)}n_{(2)a} n_{(2)b}$, we find that the extrinsic curvature projection on  $\tilde{\sigma}_{ab}$ is

\begin{eqnarray}
\label{extrinsic curvature}
\tilde{\sigma}^{ac}K_{cb}=\left(
\begin{array}{ccc}
          0 & 0 & 0 \\
           0 & 0 & 0 \\
           0 & 0 & 0
       \end{array}
       \right)\bigoplus\left(\frac{-2 f(r)+r f'(r)}{2 r\sqrt{f(r)}}\right)I_{3} \nonumber \bigoplus \left(\sqrt{f(r)}\frac{g'(r)}{2g(r)}\right) I_4
\end{eqnarray}
 and thus vanishes when $-2 f(r)+r f'(r)=g'(r)=0$. This occurs at (and near) the horizon of the $D1D5$ black hole.

 Thus from (\ref{canonical string}) we find
\begin{eqnarray}
\label{action part1}
\left[K_{tt} (x),N\sqrt{\tilde{\sigma}} U_0^{abcd}u_a n_{(1)b} n_{(1)c}u_d(\tilde{x})\right]=\hbar \delta^{D-1}(x-\tilde{x}).\\
\left[K_{yy} (x),N\sqrt{\tilde{\sigma}} U_0^{abcd}u_a n_{(2)b} n_{(2)c}u_d(\tilde{x})\right]=\hbar \delta^{D-1}(x-\tilde{x}).
\end{eqnarray}
where $K_{tt}\equiv K^{ab}n_{(1)b} n_{(1)a}$, $K_{yy}\equiv K^{ab}n_{(2)b} n_{(2)a}$ and $N=N_t=N_y$. Integrating over a closed $D-1$ hypersurface we find
\begin{eqnarray}
\label{action part2}
\left[\Theta_{r-t_E},S_{Wr-t}\right]=\hbar\\\
\left[\Theta_{r-y},S_{Wr-y}\right]=\hbar
\end{eqnarray}
where\\
 $\Theta_{r-t_E}\equiv\oint ds_1K_{tt}= \oint N dt_E K_{tt}$ is the opening angle at the $r-t_E$ surface,\\$S_{Wr-t}\equiv\oint Ndyd^{D-3}x\sqrt{\tilde{\sigma}}u_a n_{(1)b} n_{(1)c}u_d U_0^{abcd}$ (note that after using the field equation for $U_0^{abcd}$, this turns to Wald entropy \cite{Brown:1995su}) as expected,\\ $\Theta_{r-y}\equiv\oint ds_2 K_{yy}=\oint N dy K_{yy}$ is the opening angle at the $r-y$ surface,\\ $S_{Wr-y}\equiv\oint Ndt_Ed^{D-3}x\sqrt{\tilde{\sigma}}u_a n_{(2)b} n_{(2)c}u_d U_0^{abcd}$.

Finally we get a prediction of the following uncertainty relation on the horizon of a $D1D5$ black hole:
\begin{eqnarray}
\label{action part2}
\Delta \Theta_{r-t_E} \Delta S_{W}\geq \hbar\\
\Delta \Theta_{r-y} \Delta S_{Wr-y}\geq \hbar
\end{eqnarray}

The prediction of the uncertainty at the opening angle in the $r-y$ surface for $D1D5$ black hole is in agreement with the fuzzball proposal, where it was found that some specific string vibrations form metrics with conical singularities at the $r-y$ surface \cite{Mathur:2005zp,Giusto:2012yz, Lunin:2002iz, Balasubramanian:2000rt}. These metrics have different values of discrete opening angles which leads to an uncertainty at the opening angle at the $r-y$ surface.

Our derivation predicts more acceptable singularities that should come from different values of $S_{Wr-y}$. This increases the number of acceptable singularities  and may be used for finding the missing microstates of the $D1D5$ black hole \cite{Matur}.

Moreover, since this seems to come about because $t_E$ and $y$ are treated on equal footing in the Euclidean metric, this raises the idea that the uncertainty at the opening angle at the $r-t_E$ surface is not just a result of  fluctuations of a theory of quantum gravity, but may also be described as result of some specific string theory excitations. Although it seems that this kind of uncertainty comes from "Euclidian" string theory excitations which may not have a physical meaning the fact that their presence comes from canonical relations on the one hand, and their resemblance to the entropy of the black hole on the other hand, suggests that this kind of non-physical excitations may contribute to the known acceptable singularities of a $D1D5$ black hole.

 As noted at the beginning, we expect that uncertainty at the opening angle at the $r-t_E$ surface exists in ordinary spherically symmetric black holes as well. The ability to describe the same property in the $D1D5$ black hole using string theory excitations, may extend our understanding of the gravitational phase space of black holes in general.

 \textbf{Summary}

In this paper we have found that for a $D1D5$ black hole there exist in addition to the known canonical pair, $\Theta_{r-t_E}$ and $S_W$, another canonical pair: The opening angle at the $r-y$ surface, $\Theta_{r-y}$, and a Wald like term $S_{Wr-y}$. This leads to an uncertainty at $\Theta_{r-y}$ which predicts that the surfaces $r-y$ is actually a superposition of surfaces with conical singularities. This corresponds to the type of singularities obtained from string theory excitations and thus suggests that the uncertainties at $S_{Wr-y},\Theta_{r-t_E},S_W$  may also come from other type of string vibration.

\textbf{Acknowledgments:}  We thank Ramy Brustein, Sunny Itzhaki, Steve Carlip, Shmuel Elitzur, Judy Kupferman and Dan Gorbonos for valuable discussions.

\textbf{Appendix}

\textbf{Testing the condition for static spherically symmetric metric:}

For static spherically symmetric metrics,
\begin{eqnarray}
\label{Static spherical}
ds^2=-f(r)dt^2+f(r)^{-1}dr^2+r^2d\Omega ^2,
\end{eqnarray}
we choose $n^a$ to be in the time direction
\begin{eqnarray}
\label{Static spherical}
n^a=(f(r)^{-1/2},0,0,....,0).
\end{eqnarray}
$u^a$ becomes
\begin{eqnarray}
\label{Static spherical}
u^a=(0,f(r)^{1/2},0,....,0),
\end{eqnarray}
 and the extrinsic curvature projection on the $D-2$ submanifold becomes
\begin{eqnarray}
\label{extrinsic curvature}
\sigma^{ac}K_{cb}=\left(
         \begin{array}{ccccc}
            0 & 0 & 0 & \cdots & 0  \\
           0 & 0 & 0 &  & 0 \\
           0 & 0 & f^{1/2}(r)/r &  &  0 \\
           \vdots & & & \ddots  & 0 \\
           0 & 0 & \cdots & 0 & f^{1/2}(r)/r  \\
         \end{array}
       \right).
\end{eqnarray}
The extrinsic curvature projection  on the  $D-2$ surface vanishes when $f(r)=0$, namely on the horizon. (Note that since near the horizon the extrinsic curvature projection on the $D-2$ manifold is very small compared to $K_{nn}$ and can be neglected this also holds near the horizon.)
\\


\begin{thebibliography}{99}

\bibitem{Carlip:1993sa}
  S.~Carlip and C.~Teitelboim,
  Class.\ Quant.\ Grav.\  {\bf 12} (1995) 1699
  [gr-qc/9312002].

\bibitem{Wald:1993nt}
  R.~M.~Wald,
  Phys.\ Rev.\ D {\bf 48}, 3427 (1993)
  [gr-qc/9307038].

\bibitem{Iyer:1994ys}
  V.~Iyer and R.~M.~Wald,
  Phys.\ Rev.\ D {\bf 50}, 846 (1994)
  [gr-qc/9403028].

\bibitem{Brustein:2012sa}
  R.~Brustein and M.~Hadad,
  Phys.\ Lett.\ B {\bf 718}, 653 (2012)
  [arXiv:1202.5273 [hep-th]].

\bibitem{Dittrich:2008va}
  B.~Dittrich and S.~Speziale,
  New J.\ Phys.\  {\bf 10}, 083006 (2008)
  [arXiv:0802.0864 [gr-qc]].


\bibitem{Strominger:1996sh}
  A.~Strominger and C.~Vafa,
  Phys.\ Lett.\ B {\bf 379} (1996) 99
  [hep-th/9601029].


\bibitem{Mathur:2005zp}
  S.~D.~Mathur,
  Fortsch.\ Phys.\  {\bf 53}, 793 (2005)
  [hep-th/0502050].




\bibitem{Giusto:2012yz}
  S.~Giusto, O.~Lunin, S.~D.~Mathur and D.~Turton,
  JHEP {\bf 1302}, 050 (2013)
  [arXiv:1211.0306 [hep-th]]




\bibitem{Brown:1995su}
  J.~D.~Brown,
  Phys.\ Rev.\ D {\bf 52}, 7011 (1995)
  [gr-qc/9506085].

\bibitem{Avraham:2014twa}
  E.~Avraham and R.~Brustein,
  Phys.\ Rev.\ D {\bf 90} (2014) 2,  024003
  [arXiv:1401.4921 [hep-th]].

\bibitem{Lunin:2002iz}
  O.~Lunin, J.~M.~Maldacena and L.~Maoz,
  hep-th/0212210.

\bibitem{Balasubramanian:2000rt}
  V.~Balasubramanian, J.~de Boer, E.~Keski-Vakkuri and S.~F.~Ross,
  Phys.\ Rev.\ D {\bf 64}, 064011 (2001)
  [hep-th/0011217].

\bibitem{Matur}
   S.~D.~Mathur,
   ``Confusions and questions about the information paradox'',
    http://www.physics.ohio-state.edu/\~~mathur/confusions2.pdf.


\end{thebibliography}
\end{document}